\documentclass{JHEP3}

\input{epsf}
\usepackage{epsfig}
\usepackage{amssymb}
\usepackage{amsfonts}
\usepackage{amsbsy}

\def\IIb{{IIb}}
\def\susy{{susy}}

\def\rem{$\clubsuit$}
\def\re{\mbox{Re }}

\def\IZ{\mathbb{Z}}
\def\IR{\mathbb{R}}

\def\CK {{\cal K}}
\def\CN {{\cal N}}
\def\CR {{\cal R}}

\def\CF {{\cal F}}

\def\CL {{\cal L}}

\def\CK{{\cal K}}

\def\half{\frac{1}{2}}

\newcommand{\eq}[1]{Eq.~(\ref{eq:#1})}

\renewcommand{\Im}{{\rm Im }}
\renewcommand{\Re}{{\rm Re }}

\def\one{{\hbox{ 1\kern-.8mm l}}}

\def\tr{{\rm tr\,}}

\def\p{\partial}
\def\ba{\bar{a}}
\def\bb{\bar{b}}

\def\be{\bar{e}}
\def\bz{\bar{z}}
\def\bZ{\bar{Z}}
\def\bW{\bar{W}}
\def\bD{\bar{D}}

\def\bpartial{\bar{\partial}}

\def\bpsi{\bar{\psi}}
\def\bF{\bar{F}}

\def\bz{\bar{z}}

\def\bZ{\bar{Z}}
\def\GeV{{\rm GeV}}
\def\TeV{{\rm TeV}}

\title{Distributions of nonsupersymmetric flux vacua}

\author{Frederik Denef$^{1}$ and Michael R. Douglas$^{1,\&}$\\
$^1$NHETC and Department of Physics and Astronomy\\
Rutgers University\\
Piscataway, NJ 08855--0849, USA\\
\\
$^\&$I.H.E.S., Le Bois-Marie, Bures-sur-Yvette, 91440 France\\
{\tt denef, mrd@physics.rutgers.edu}
}

\abstract{
We continue the study of the distribution of
nonsupersymmetric flux vacua in \IIb\ string theory
compactified on Calabi-Yau manifolds, as in hep-th/0404116.

We show that the basic structure of this problem is that of finding
eigenvectors of the matrix of second derivatives of the superpotential,
and that many features of the results are determined by features
of the generic
ensemble of such matrices, the CI ensemble of Altland and
Zirnbauer originating in mesoscopic physics.

We study some simple examples in detail, exhibiting various factors
which can favor low or high scale supersymmetry breaking.
}

\begin{document}

\section{Introduction}

In this work we continue the study of the statistics of flux vacua
initiated in \cite{stat,ad,DSZ}, and continue the study of
supersymmetry breaking vacua begun in \cite{dd}.  For
a general introduction to this problem and its applications,
and further references, see \cite{basic}.  Some more recent works
on vacuum statistics and related topics include
\cite{Conlon:2004ds,Dine:2004ct,Dvali:2004tm,Kumar:2004pv,Blum}.

Our main result will be explicit formulae for the density of
supersymmetry breaking vacua in an ensemble of effective supergravity
field theories (EFT's) in which the superpotential is a ``random
variable'' in a sense we define below.  This type of ensemble includes
the set of EFT's obtained by compactifying string/M theory with
fluxes, in the limit that flux quantization can be neglected.  We will
give specific results for type \IIb\ flux compactification on
Calabi-Yau \cite{GKP}, which at present is the best studied (and most
calculable) example, but stress that the techniques (and possibly many
features of the results) are more general.

Although we foresee many applications of such results to string
duality and phenomenology, a question of primary current interest is
the likelihood to discover supersymmetry at the energy range $1 \TeV$
to be probed at LHC, perhaps as the traditional MSSM scenarios,
``split supersymmetry'' \cite{ArkDim,GR}, or perhaps other scenarios.  As
discussed in \cite{stat,basic}, a systematic way to study this
question is to try to count or estimate the number of string vacua
which realize the various scenarios.  Even with our limited present
understanding, it is conceivable that we can argue that some scenarios
are so unlikely that we should regard their observation as evidence
against string theory, in other words we would have derived a
strong prediction from string theory.  Admittedly, this possibility
depends on optimistic assumptions about the total number of string
theory vacua, but in any case we believe vacuum counting will give
essential information for any approach to string phenomenology.

The main reason to reconsider traditional attitudes towards
naturalness and supersymmetry is that, as becomes clear from a
quantitative approach, the advantage gained by low energy
supersymmetry is just not that large, compared to the current
estimates of the number of string theory vacua; $N_{vac} >> 10^{100}$
vacua, based both on estimates of the number of flux vacua
\cite{BouPol,ad} and on the idea that this large number of vacua is
actually needed to solve the cosmological constant problem
\cite{BT,Weinberg,BouPol}.

To quantify the study of supersymmetry breaking, we adopt the following
simplified description of the problem.  We need to
find the joint distribution of the Higgs mass,
supersymmetry breaking scale and cosmological constant,
$$
d\mu[M_H,M_{\susy},\Lambda]
$$
among ``otherwise phenomenologically acceptable'' vacua, where
\begin{equation}\label{eq:defMsusy}
M_{\susy}^4 = \sum_A |F_A|^2 + \sum_\alpha D_\alpha^2 .
\end{equation}
We then evaluate this at the (presumed) values $M_H \sim 100 \GeV
\sim 10^{-17} M_{planck}$ and $\Lambda \sim 10^{-122} M_{planck}^4$,
and look at the resulting distribution of supersymmetry breaking
scales.  Rough arguments \cite{mrd-susy,suss-susy,dgt} summarized
in \cite{basic} suggest that for $M_H^2 \ge M_{\susy} M_{planck}$,
this goes as
\begin{equation} \label{eq:hierarchy}
d\mu[M_H,M_{\susy},\Lambda] \sim
\frac{dM_H^2}{M_{\susy}^4 / M_{planck}^2}
\frac{d\Lambda}{M_{planck}^4} d\mu[M_{\susy}] .
\end{equation}
To get a sense of what this means, consider a simple power law ansatz,
$$
d\mu[M_{\susy}] \sim M_{\susy}^{\alpha} dM_{\susy}^2
$$
Given such a distribution, the distribution of physical vacua would
be weighed towards high scales if $\alpha>2$, showing that
it would not take a huge growth in the number of vacua with breaking
scale to dominate the advantage in solving the hierarchy problem.
\cite{suss-susy,mrd-susy}
Another way to make this point is to note that the ratio
$(M_{susy\ high}/M_{susy\ low})^4$
for two plausible guesses at the supersymmetry breaking scale, the
high scale $M_{susy \ high} \sim 10^{16}\GeV$
and the intermediate scale $M_{susy \ low} \sim 10^{10}\GeV$,
is $10^{24}$.  This is not a large number in the present context.

Of course the true distribution of supersymmetry breaking scales in
string/M theory vacua will be far more complicated than this simple
power law ansatz, with various components reflecting various
supersymmetry breaking mechanisms and other structure in the problem.
While it will clearly take a great deal of work to form any real
understanding of this true distribution, we will see in the following
that certain generic features of the problem, following just from the
the structure of supergravity and of generic distributions of
EFT's, do translate into specific properties and factors in the
distributions, which might well be shared by the true distribution.
Furthermore, many aspects of string compactification which at first
sight look highly significant, such as mechanisms to produce
hierarchically small scales, can turn out to be no more important than
these generic properties in the final statistics.

We turn to detailed considerations of the problem of counting pure
F breaking vacua in sections 2 and 3.  Starting from a precise
ensemble of EFT's such as those obtained by flux compactification,
this is a mathematically well defined problem, but not a simple one.
While we will obtain precise formulae as in \cite{ad,dd}, to some
extent a more useful description of the basic result is that the
distribution of supersymmetry breaking vacua is similar to that of
supersymmetric vacua, with certain ``correction factors'' which we
will explain in detail.  We also discuss the likely effects of
taking flux quantization into account.

In section 4 we make some comments on the distribution obtained by
supersymmetry breaking by antibranes, or more generally by D terms.

Section 5 contains conclusions.

\section{Preliminary considerations}

Our starting point is the $\CN=1$ supergravity  potential and its
derivatives,\footnote{
The covariant derivatives $D_a$ are both
K\"ahler and metric covariant, {\it i.e.} when acting
on $W$ and its derivatives include the K\"ahler
connection $\partial_a \CK$, and when acting on tensors
include the Levi-Civita connection $\Gamma^b_{ac}$.
$R$ is the curvature of the cotangent bundle,
i.e. ${R^d}_{ca \bb} \, X_d \equiv [\nabla_a,\bar{\nabla}_{\bb}] X_c
= \bpartial_{\bb}(g^{\be d}
\partial_a g_{c \be}) \, X_d$. At a vacuum $dV=0$, it does not matter whether the outer derivatives
are covariant derivatives or ordinary partial derivatives, because
$D d V = d^2 V$.  We set $M_p=1$.}
\begin{eqnarray}  \label{eq:defV}
V &=& e^{{\CK}}\left( g^{a\bb} D_aW \bD_{\bb} \bW -
 {3} |W|^2 \right) + D^2 \\
\label{eq:Vderivs}
 \partial_a V &=& e^{\CK} ( D_a D_b W \bD^b \bW - 2 D_a W \bW )
 \label{eq:DV} \\
 D_a \partial_b V &=& e^{\CK} (D_a D_b D_c W \bD^c \bW - D_a D_b W \bW)
\label{eq:DDV}\\
 \bD_{\ba} \partial_b V &=&
 e^{\CK}({R^d}_{c \ba b} D_d W \bD^c \bW
 + g_{b\ba} D_c W \bD^c \bW - D_b W D_{\ba} \bW \nonumber \\
  && - 2 g_{b\ba} W \bW
 + D_b D_c W \bD_{\ba} \bD^c \bW) . \label{eq:DcDV}
\end{eqnarray}
For simplicity we will first assume the D terms are zero or at least
independent of the fields (until section 4).

The approach we will take to finding statistics of vacua
is to summarize the sum over all choices which go into the
potential \eq{defV}, in terms of a
joint distribution $d\mu$ of the superpotential $W$ and its derivatives,
evaluated at a point $z$ in the space of chiral field vevs, a candidate
vacuum.
To evaluate all of Eqs. (\ref{eq:Vderivs})--(\ref{eq:DcDV}) at a point $z$,
the joint distribution must describe
$W(z)$, $\CK(z,\bz)$, and up to three
derivatives of $W$ at $z$.

The explicit dependence of these quantities on $\CK(z)$, the
K\"ahler potential at $z$, can be removed by either making
redefinitions such as $\hat W = e^{\CK/2} W$ (as in \cite{dd}), or
equivalently making a K\"ahler-Weyl transformation to set
$\CK(z,\bz)=0$.
This does not change any of the equations
because $e^{\CK}$ is covariantly constant with respect to $D_a$.
Let us now assume this has been done, and use the notations
\begin{equation}
F_A = D_A W(z); \qquad Z_{AB} = D_A D_B W(z); \qquad U_{ABC} = D_A
D_B D_C W(z)
\end{equation}
where
capital indices $A,B$ etc.
are defined to be orthonormal complex indices with
respect to the K\"ahler metric at $z$.
We retain the symbol $W$ for $W(z)$.
The tensors $Z$ and $U$ are symmetric in all indices.

We can then rewrite Eqs. (\ref{eq:Vderivs})--(\ref{eq:DcDV}) as
\begin{eqnarray} \label{eq:dV}
 \partial_A V &=& Z_{AB} \bF^{B} - 2 F_A \bW \\
 D_A \partial_B V &=& U_{ABC} \bF^K - Z_{AB} \bW \\
\bar D_A \partial_B V &=& R_{ABCD} F^C \bF^D + \delta_{AB} |F|^2
 - \bF_A F_B -2 \delta_{AB} |W|^2 + \bZ_{AC} Z_{BD} .
\end{eqnarray}
Given a joint distribution
\begin{equation}\label{eq:dist}
 d\mu[W, F_A, Z_{AB}, U_{ABC}] ,
\end{equation}
such as might come from summing over all choices of flux in a given
compactification, we will then compute densities such as
$$
\rho(z) = \int d\mu\ \delta(V'(z)) |\det V''(z)| \theta(V'')
$$
in terms of the joint distribution.

\subsection{Distributions}

The joint distribution \eq{dist}
for an integral over fluxes in \IIb\ on CY in the large volume limit
was worked out in \cite{dd}.  To summarize, in these effective theories
the K\"ahler potential $\CK$ is independent of the flux, while
the GTVW formula
\begin{equation}\label{eq:gtvw}
 W = \int G \wedge \Omega(z)
\end{equation}
tells us that the superpotential is linear
in the flux $G$.  Thus, $W(z)$ and
all of its derivatives are linear in the flux, and one can
change variables in the integral over fluxes to
find a simple joint distribution in which $W$, $F_0$, $F_I$
and $Z_{0I}$ are independent. Here the index `0' refers to the
dilaton, and $I=1,\ldots,n$ to the complex structure moduli.
Using threefold special geometry, for the particular case of
\IIb\ flux vacua, the variables $Z_{AB}$ and $U_{ABC}$ are
determined in terms of $Z_I \equiv Z_{0I}$ and $F_A$; for example
 $$
Z_{IJ} = \CF_{IJK} \bZ^K .
$$
Thus, the resulting distribution is
\begin{equation} \label{eq:IIbdist}
d\mu_{IIb} = d^2W~ d^{2n+2}F_A~ d^{2n}Z_{0I}~ \delta(L-|W|^2+|F|^2-|Z_{0I}|^2)
\end{equation}
with the other variables determined in terms of these.
One could also express the distribution for the original problem with
quantized fluxes in this way, as a sum over lattice points embedded in
the parameter space in a way determined by $z$ and the periods.

The superpotential \eq{gtvw} does not include K\"ahler moduli and
as in \cite{dd} we will ignore these moduli. Of course, in an exact
description, these must be taken into account. In particular,
their stabilization requires a sufficient number of nonperturbative
contributions to the superpotential \cite{KKLT,ddf,ddfgk}, which may
or may not be present in a given model. Moreover, to stabilize the
volume at a reasonably large value, moderately small values of
$W_{\rm flux}$ are needed. However, if thus stabilized, the K\"ahler
sector has relatively little influence on the statistics and
properties of vacua, because its contribution to the potential is
exponentially small, and because by far the main degeneracy of vacua
comes from the different choices of flux. In particular, inclusion
of K\"ahler moduli typically only produces small shifts in the vacua
found by ignoring them. This was made more precise in \cite{ddf}
section 4.1, and \cite{Dine:2004ct}. The constraint of small $W_{\rm
flux}$, needed to stabilize the model in a controlled (large radius)
regime, merely reduces the number of suitable vacua by a factor
$|W_{\rm flux}|^2$ \cite{dd}, and anyway this condition will be met
automatically for the phenomenologically most relevant vacua we will
study in the following, namely those with $\Lambda \sim 0$ and
$M^2_{\rm susy} \ll 1$.

The requirement of a sufficient number of nonperturbative
contributions is more subtle, but the results of \cite{ddf,ddfgk}
suggest that many models satisfying this should exist, although
completely explicit constructions tend to be computationally complex. We
suspect that major progress on this point would come from developing
other ways of understanding K\"ahler moduli stabilization. For
example, mirror symmetry suggests the existence of type IIB
non-Calabi-Yau deformations dual to turning on IIA NS-NS flux, which
should thus be described by a similar superpotential depending on the
IIB K\"ahler moduli. In this case, the K\"ahler sector would also
contribute significantly to the counting of vacua, but since it would
be governed by a flux type superpotential, it would also fit into the
general class of supergravity ensembles for which the analysis below
should be valid.  We discuss this point a bit further in section 3.2.2.

In any case, this discussion, together with the fact that IIB flux
vacua are computationally very accessible due to the underlying
special geometry structure, justifies the claim that such ensembles
are good models for the statistics of string vacua.

One could do a similar computation for any class of models in
which $\CK$ and $W$ are computable.
Generically, one would expect to obtain a similar distribution, in
which a sum over $K$ flux (or other) parameters leads to a rough
independence for the first $K$ quantities in the distribution.  In
theories (such as the heterotic string) with fewer fluxes than the
\IIb\ string, this would lead to constraints between the $Z$'s and the
$(W,F)$ variables, while in theories with more parameters (such as F
theory on fourfolds) one has more parameters and thus might expect
a more generic ensemble of matrices $Z_{IJ}$.

While it remains to be seen which distribution best represents the
full set of string/M theory vacua, it seems plausible that this would
be the one with the most free parameters, {\it i.e.} the fourfolds.
One furthermore expects quantum corrections to these classical flux
superpotentials.  Thus another interesting ensemble, which might
usefully represent this more generic distribution, is simply to
take all of the parameters to be independent,
\begin{equation}\label{eq:gendist}
d\mu_{G} = d^2W d^{2m}F_A d^{m(m+1)}Z_{AB} \delta(L-|W|^2+|F|^2-|Z|^2)
\end{equation}
where the delta function is an ansatz for a cutoff analogous to the
one which made the number of \IIb\ flux vacua finite.  This will turn
out to be related to distributions previously considered in random
matrix theory, as we discuss in section 2.4.

To complete the specification of such a model ensemble, one must
define the variables $U_{ABC}$.  Now in the fourfolds and other
examples we might model this on, the $U$ variables would still be
determined in terms of the others, which might be important.  One
might also argue that a better model ensemble would be a simple
proposal which keeps more of the structure of Calabi-Yau moduli
spaces and their degenerations.  These are interesting ideas for
future work, but as we will argue below \eq{gendist} already shares
interesting features with more realistic distributions.

In any case, we will keep most of the discussion general, and not
assume any {\it a priori} relation between the variables in \eq{dist}.

\subsection{Solving the equations $V'=0$}

The equations $d V = 0$ are quadratic in the parameters,
and the best way to exhibit
their structure is to rewrite them as a matrix $N(W,Z)$ depending
on $(W,\bar W,Z,\bar Z)$ acting on the vector $F_A$, as
\begin{eqnarray} \label{eq:defN}
0 = \left(\begin{array}{c}\p_A V\\ \bpartial_A V
\end{array}\right) &=&
 \left(\begin{array}{cccc}
  -2\bW & Z_{AB} \\
  \bZ_{AB} & -2W
 \end{array} \right)
\left(\begin{array}{c}F^B\\ \bF^B
\end{array}\right) \\
&=&  \left(\begin{array}{cccc}
  -2|W| & e^{-i\theta} Z_{AB} \\
e^{i\theta} \bZ_{AB} & -2|W|
 \end{array} \right)
\left(\begin{array}{c}e^{-i\theta} F^B\\ e^{i\theta}\bF^B
\end{array}\right)
\end{eqnarray}
where $e^{i\theta}=W/|W|$.

Define the matrix $N(W,Z)$ to be the
second of these matrices ({\it i.e.} the matrix depending on $|W|$),
and write
$$
N = M -2|W|\cdot{\bf 1}
$$
in terms of the matrix
\begin{equation}\label{eq:deltaH}
M =
 \left(\begin{array}{cccc}
  0 & e^{-i\theta} Z_{AB} \\
e^{i\theta} \bZ_{AB} & 0
 \end{array} \right) .
\end{equation}
Since $Z_{AB}$ is symmetric,
the matrix $M$ is hermitian, so it has an orthonormal
basis of eigenvectors.  Thus, we see that
non-supersymmetric vacua (solutions of $V'=0$ with $F\ne 0$)
correspond to eigenvectors of the matrix $M$
with eigenvalue $2|W|$.

Clearly the matrix $M$ is not a generic hermitian matrix;
it has additional symmetry properties.
First, its eigenvectors come in pairs with opposite
eigenvalues $(+\lambda_a,-\lambda_a)$.
These eigenvalues
are independent of $\theta$; this follows because their squares
are the eigenvalues of the hermitian matrix $Z\bar Z$. The
corresponding eigenvectors are
\begin{equation}\label{eq:genpsi}
\Psi_a^+
 = \left(\begin{array}{c} e^{-i\theta/2}\psi_a\\
e^{i\theta/2} \bpsi_a
\end{array}\right), \qquad
\Psi_a^-
 = \left(\begin{array}{c} i e^{-i\theta/2}\psi_a\\
- i e^{i\theta/2} \bpsi_a
\end{array}\right),
\end{equation}
where $\psi_a$ solves
\begin{equation} \label{eq:takagidecom}
 Z \bpsi_a = \lambda_a \psi_a
\end{equation}
and we take $\lambda_a \geq 0$. Because the $\psi_a$ are
eigenvectors of the hermitian matrix $Z \bar Z$, we can take them to
be orthonormal. Then, defining the unitary matrix $U$ by $U_{Ab}
\equiv \bpsi_{b,A}$, we have
\begin{equation} \label{eq:zdiag}
  Z = U \lambda U^T
\end{equation}
where $\lambda = \mbox{diag}(\lambda_a)$. Any symmetric complex
matrix can be decomposed in this way.

Thus, the generic supersymmetry-breaking solutions to (\ref{eq:defN})
can be written as
\begin{equation} \label{eq:solconstr}
  2 W = \lambda_a e^{i \theta}; \qquad F = f e^{i \theta/2} \psi_a
\end{equation}
with $f \in \IR$. Varying $\theta$ in $[0,2\pi]$, this fills out a
one complex dimensional subspace.

\subsection{Masses of moduli} \label{sec:masses}

We have just seen that non-supersymmetric solutions of $V'=0$
correspond to eigenvectors of a matrix $M$ with eigenvalue
$2|W|$.  Thus, in ensembles such as the \IIb\ flux
ensemble in which the parameters $W$ and $F_A$
can be freely varied, such vacua are {\bf generic}.

Of course, the vacua of most interest are metastable, {\it i.e.}
the bosonic mass matrix $V''$ has no negative eigenvalues.
This mass matrix is
\begin{equation}\label{eq:defdtwoV}
d^2V = (M + |W|)(M -2|W|) + V''_1 + V''_2
\end{equation}
where the successive terms are of order $F^0$, $F^1$ and $F^2$.

The $F^0$ term is the product of the matrix $N$ above with the
matrix
 $$ H = M + |W| \cdot {\bf 1} $$
which is the same as the matrix $d^2|W|$ whose determinant enters
into the density of supersymmetric vacua \cite{dd}.
The higher order terms are
 $$
 V''_1 = \left(\begin{array}{cccc}
  0 & S_1 \\ \bar{S}_1 & 0
 \end{array} \right),
 \qquad S_1 = U_{ABC} \bF^C
 $$
and
 $$
 V''_2 =
 \left(\begin{array}{cccc}
  S_2 & 0 \\ 0 & \bar{S}_2
 \end{array} \right), \qquad
 S_2 = R_{A \bar{B} C \bar{D}} \bF^C F^D + \delta_{AB} |F|^2 - F_A \bF_B.
 $$

We now assume $|F|<<M_p\equiv 1$.
In this case, the eigenvectors $\Psi_a^{\pm}$ of $M$
defined in (\ref{eq:genpsi}), are approximate eigenvectors of
$d^2V$. Let us take the eigenvalues $\lambda_A$ to be ordered,
\begin{equation}\label{eq:ordering}
0 \le \lambda_1 \le \lambda_2 \le \cdots \le \lambda_m .
\end{equation}
The values of $V''$ in the $\Psi_a$ directions are
\begin{eqnarray}
 (m_a^{\pm})^2 &=& \langle \Psi_a^{\pm},d^2V \Psi_a^{\pm} \rangle
 \nonumber \\
 &=& (\pm \lambda_a + |W|)(\pm \lambda_a - 2|W|) \pm 2\,
   \re(e^{i \theta} \bpsi_a S_1 \bpsi_a)
 + 2 \, \bpsi_a S_2 \psi_a. \label{eq:masses}
\end{eqnarray}

By (\ref{eq:defN}), one of these eigenvectors, call it $\Psi^+_F$,
is proportional to
$\left(\begin{array}{cc}e^{-i\theta} F^B& e^{i\theta}\bF^B
\end{array}\right)$, with $\lambda = 2 |W|$.  There will also
be a complementary eigenvector $\Psi^-_F$ with eigenvalue $-2|W|$.

Let us first consider variations of the moduli proportional to the
other eigenvectors.  Suppose the corresponding eigenvalue $\lambda_a$
is greater than $2|W|$. In this case, the first term in
(\ref{eq:masses}) will be positive, and if the gap $\lambda_a-2|W|$ is
sufficiently large (as will be the case for generic $Z$ when $F$ is
small), this term dominates the higher order terms.  On the other
hand, positive eigenvalues less than $2|W|$ will produce a negative
$m^2$, unless they are very close to $2|W|$ and the higher order terms
are fine tuned to compensate for the negative lowest order
term. Thus, the bulk of tachyon-free nonsupersymmetric vacua with
sufficiently small $F$ will  come from solutions for which
$2|W|$ equals the {\bf lowest} eigenvalue $\lambda_1$.

We now consider $(m_1^{\pm})^2 = (m_F^{\pm})^2$.  For $m_1^+$, the
first term in \eq{masses} vanishes, so
$V''$ in this direction is given by the
matrix element of the higher order terms:
\begin{equation} \label{eq:msq}
(m_F^+)^2 = \frac{2}{|F|^2} \left(\Re ( e^{2i\theta} U_{ABC} \bF^A
\bF^B \bF^C ) + R_{A\bar{B}C\bar{D}} \bF^A F^B \bF^C F^D \right).
\end{equation}
Similarly, the matrix element for the complementary eigenvector with
eigenvalue $-2|W|$ is
\begin{equation} \label{eq:msqn}
(m_F^-)^2 = 4 |W|^2 + \frac{2}{|F|^2} \left(
 - \Re ( e^{2i\theta} U_{ABC} \bF^A \bF^B \bF^C )
 + R_{A\bar{B}C\bar{D}} \bF^A F^B \bF^C F^D \right).
\end{equation}

There are now two cases to distinguish.  The simplest case
is $|F|<<|W|$, but this can only be compatible with
$V=|F|^2-3|W|^2+|D|^2\sim 0$ if $|D|\sim |W|$, {\it i.e.}
the supersymmetry breaking is dominated by the D terms.
If we have $D=0$ or even $|D|\sim |F|$, then $|F|\sim |W|$ and
we must take this into account.

In the first case, and for generic $U_{ABC}$, the first term in
\eq{msq} will dominate, so $(m_F^+)^2$ will be positive for half the
integration domain of $\theta$. Furthermore, if $|W|$ is not too
small (greater than $O(\sqrt{F})$), the first term in \eq{msqn} will
dominate and $(m_F^-)^2$ will also be positive. Hence the
requirement of metastability puts only a mild constraint on the
integration domain in this case.

On the other hand, if $|F| \sim |W|$, the first term in \eq{msqn} is
of order $F^2$ and no longer dominates. Instead, for generic $U$ and
$\theta$, either $(m_F^+)^2<0$ or $(m_F^-)^2<0$, as the $O(F)$ term
appears with different signs in these. Thus, to get a metastable
vacuum with zero (or positive) cosmological constant, we have to
fine-tune the $O(F)$ term such that it becomes smaller than the
$O(F^2)$ term:
\begin{equation} \label{eq:premetaconst}
 \Re ( e^{i\theta/2} U_{ABC} \bpsi^A \bpsi^B \bpsi^C ) <
 O(F).
\end{equation}
This can be achieved by tuning $\theta$ in an interval of size
$\sim |F|$, which, if the measure for $\theta$ is otherwise
uniform, will give an additional suppression of the expected
number of vacua by a factor of $|F|$.


However, because the diagonal mass matrix elements are now tuned one
order in $F$ smaller, we have to be a bit more careful and also
consider the off diagonal matrix element between $\Psi_1^+$ and
$\Psi_1^-$.\footnote{We thank Michael Dine, Deva O'Neil and Zheng
Sun for pointing this out to us.} This is
\begin{equation}
 \langle \Psi_1^-,d^2V \, \Psi_1^+\rangle =
 2\, \Im(e^{i \theta} \bpsi_1 S_1 \bpsi_1) = \frac{2}{|F|^2}
 \Im ( e^{2i\theta} U_{ABC} \bF^A \bF^B \bF^C ).
\end{equation}
Denoting $s_1 \equiv e^{i \theta} \bpsi_1 S_1 \bpsi_1$, $s_2 \equiv
\bpsi_1 S_2 \psi_1$, the determinant of the $2 \times 2$ mass matrix
in the 1-directions is thus
\begin{equation}
 \det = 4\left(|W|^2(\Re s_1 + s_2) - |s_1|^2 + s_2^2\right).
\end{equation}
Note that if $W \sim F$, the leading term in the expansion in powers
of $F$ is generically $-|s_1|^2 \sim F^2$, and this is always
negative \emph{independent} of $\theta$. Therefore, to avoid a
tachyon, we really need $|s_1| < O(F)^2$, so (\ref{eq:premetaconst})
gets replaced by
\begin{equation} \label{eq:metaconst}
 | U_{ABC} \bpsi_1^A \bpsi_1^B \bpsi_1^C | < O(F).
\end{equation}
Assuming approximate uniform distribution of this complex component
of $U$, this metastability constraint will therefore give an
additional $O(F^2)$ suppression rather than the $O(F)$ we deduced
neglecting the off-diagonal element. The masses $m_1^{\pm}$ on the
other hand will still be of order $F$.


In any case, the overall conclusion of this analysis is that
metastability is a relatively mild constraint, in the sense that it
does not drastically reduce the number of non-supersymmetric vacua.
In general, we find that a rough fraction $1/n$ of all
nonsupersymmetric critical points are metastable vacua (since we can
only use one of the $n$ eigenvalues). This is much larger than the
naive estimate of $2^{-2n}$, obtained by taking the $m_i^2$
independent and symmetrically distributed around zero. This is
consistent with the intuition that, for $|F| << |W| << 1$, the
situation is similar to that for global supersymmetry, in which
stability is automatic.  But the arguments here apply far more
generally than in this limit.

Nevertheless, metastability can significantly influence the
distribution: the distribution of the lowest eigenvalue can
be different from that of an arbitrary eigenvalue, and constraints
such as \eq{metaconst} are important.

\subsection{Degeneracies and the dimension of the solution space}
\label{sec:degeneracies}

We have just seen that for generic parameters $W$ and $Z$, the
space of possible supersymmetry breaking parameters $F$ is one
complex dimensional, the eigenvector of $M$ with
the lowest positive eigenvalue,
multiplied by a general phase factor.

However, at special loci in $Z$ space, the matrix $M$ might
have $k$-fold degenerate eigenvalues. In this case, the vector $F$
will vary in a $k$ real dimensional space, and varying $\theta$ will
fill out a $k+1$ real dimensional subspace.  This is the structure
which might lead to power law growth of the number of vacua with the
supersymmetry breaking scale, as suggested in
\cite{suss-susy,mrd-susy}.

An explicit example in which this is realized is the
``anti-supersymmetric branch'' of \IIb\ flux vacua, with $Z=W=0$ or
equivalently imaginary anti-self-dual flux. Now these vacua are
physically not interesting, because they have cosmological constant at
least of the order of the string scale \cite{dd}, and moreover it can
be shown that they always have a tachyonic mode.  But the number of
these vacua does grow as a high power of $F$, so we should not
immediately dismiss the idea that some physically sensible subset of
the vacua behaves in the same way.

However, if we ask what we need to get degenerate eigenvalues in other
circumstances, we find that they are non-generic, in the sense that one
must tune more than $k-1$ parameters to get a $k$-fold eigenvalue
degeneracy.  Because of this, it will turn out that these ``higher
branches'' of non-supersymmetric vacua are of lower dimension than the
primary branch, and thus will not contribute to the overall volume
estimate, and thus to the leading large $L$ asymptotics for the number
of vacua.

The arguments are analogous to the more familiar discussion
of degenerate eigenvalues in families of hermitian matrices,
so let us review this first.  For hermitian matrices,
one must tune $k^2-1$ real parameters to get a $k$-fold degeneracy.  The
simplest way to see this is to consider the change of variables to
eigenvalues and eigenvectors $M = U^\dag \lambda U$.  This is
generically unambiguous up to permutation of eigenvalues, and up to a
$U(1)^n$ left action on $U$.  Thus the $n^2$ real parameters of $M$ go
over to $n$ eigenvalues and $n^2-n$ coordinates on $U(n)/U(1)^n$.
However, a diagonal matrix with a $k$-fold degenerate eigenvalue is
preserved by conjugation by an element of $U(k)$, and thus such
matrices form a stratum in the larger space of all matrices of
real codimension $k-1$ (required to tune the eigenvalues) plus
$k^2-k$ (the dimension of the stabilizer group).

To make the analogous argument for the case at hand, we
consider the change of variables \eq{zdiag}.  Now the $n(n+1)$
real parameters in $Z$ are rewritten as $n$ eigenvalues and $n^2$
parameters of the $U(n)$ group element $V$.  On the other hand, a
$k$-fold degenerate diagonal matrix will be preserved by a unitary
$V$ satisfying $V V^T = 1$, in other words an element of $SO(k)$.
Thus the degenerate matrices form a stratum of total
real codimension $(k+2)(k-1)/2$, the dimension $k(k-1)/2$ of $SO(k)$
plus $k-1$ for the eigenvalues.

A related way to see this
is to consider the change of variables from the Lebesgue
measure on matrix elements, to eigenvalues and unitary group elements.
For the hermitian matrix, this is
$$
d^{N^2}M = [dU] \prod_i d\lambda_i \prod_{i<j} (\lambda_i-\lambda_j)^2
$$
where $[dU]$ is the invariant measure on $U(n)$.
This exhibits the extra real codimension $k(k-1)$ as an explicit
scaling dimension in the Jacobian, in analogy to the
scaling $r^{D-1} dr$ of Lebesgue measure in polar coordinates.

The analogous matrix ensemble for the present case is determined by
the symmetry properties of \eq{deltaH}.
Introducing a triplet of
Pauli matrices $\sigma_i$ to act on the $2\times 2$
block structure of \eq{deltaH}, these are
\begin{equation}\label{eq:CIdef}
M = M^\dag = - \sigma_3 M \sigma_3
 = - \sigma_2 M^* \sigma_2 .
\end{equation}
These symmetry properties are formally equivalent
to those of the ``CI ensemble'' of Altland and Zirnbauer \cite{AltZ}.
This was introduced as an ensemble of Hamiltonians of
mesoscopic systems, and in that context these symmetry
properties are time reversal invariance and spin rotation invariance.
While the physics and the interpretation of the
symmetry properties is different here, and the actual ensembles arising
from string theory (as discussed in section 2.1) are a subset of
the CI ensemble, certain properties of the CI ensemble and
specifically the structure of eigenvalue degeneracies should be shared
by the stringy ensembles.

Thus, we consider the CI ensemble measure, in which the different
matrix elements (consistent with \eq{CIdef})
are independently distributed; this is just the ensemble \eq{gendist}.
We then perform the change of variables
 $$ M = U^\dag
 \left(\begin{array}{cccc}
  \lambda& 0\\ 0& -\lambda
 \end{array} \right) U .
 $$
The CI measure transforms to
\begin{equation} \label{eq:CIdist}
 d\mu[\lambda,U] = [dU] \prod_a
 d(\lambda_a^2) \prod_{a<b} |\lambda_a^2-\lambda_b^2|
\end{equation}
and we again exhibit the codimension $k(k-1)/2+(k-1)$ of the
symmetry locus as a scaling dimension, as well as the fact that
the true invariants of $M$ are the squares of the eigenvalues.

Now, given a set of $k$ eigenvectors $\psi_a$, $1\le a\le k$, all with
eigenvalue $\lambda$, the analog of \eq{solconstr} is
\begin{equation} \label{eq:solconstrtwo}
  2 W = \lambda e^{i \theta}; \qquad F = f^a e^{i \theta/2} \psi_a
\end{equation}
where $f^a$ is a real $k$ component vector.
Thus, while the associated locus of
nonsupersymmetric vacua is $k+1$ real dimensional, and the
integral over fluxes $d^{2n}F$ would lead to an additional power law growth
$|F|^{k-1}$ (compared to the case $k=1$), the variables $Z_{IJ}$
must satisfy $(k+2)(k-1)/2$ additional real constraints, leading
to a total real codimension $(k+2)(k-1)/2-(k-1)=k(k-1)/2$ for the
$k$-fold degenerate branch. Thus already for $k=2$ these branches are
of lower dimension in the generic case, and will have volume zero.

A similar discussion applies in the special case of a $k$-fold
degenerate zero eigenvalue, but now both $\Psi^+$ and $\Psi^-$ can
be superposed in \eq{solconstr}, so the resulting locus of
nonsupersymmetric vacua is $2k$ real dimensional.  On the other
hand, a $k$-fold degeneracy at zero generically appears in real
codimension $2(k-1)+k(k-1)=(k+2)(k-1)$, so again the resulting
branch is of lower dimension.

There is a loophole in these arguments, which is the assumption of
genericity.  If the matrix $Z_{IJ}$ depends on the fluxes and fields in a
non-generic way, such that a $k$-fold degeneracy appears with real
codimension $k-1$, then the corresponding branch of vacua would
have the same dimension as the primary branch.  This is what happens
in the ``anti-supersymmetric'' branch: because of the relation
$Z_{IJ}=\CF_{IJK}\bar Z^K$, one need tune only the $n$ complex parameters
$Z_{0I}$ to obtain $Z=0$ and a $2n$-fold degeneracy at zero, so this
branch is again of the full dimension.  Since the matrices $Z_{IJ}$
obtained in \IIb\ flux compactification are definitely not generic,
we need to understand this point.

A familiar way to get a $k$-fold eigenvalue degeneracy in codimension
$k-1$ for a family of hermitian matrices is to take
the sum of $k$ {\bf commuting} matrices,
\begin{equation} \label{eq:linfam}
M = \sum t_i M_i \ {\rm with}\ [M_i,M_j]=0.
\end{equation}
In this case, since can simultaneously diagonalize the $M_i$, the
problem reduces to considering families of eigenvalues, for which there
is no difficulty in tuning to degeneracies.
Conversely, if the family is a linear sum \eq{linfam}, this is
the only way to get eigenvalue degeneracies; if some $[M_i,M_j]\ne 0$
then eigenvalue repulsion always makes the codimension
of a $k$-fold degeneracy (reachable
by tuning parameters which include $t_i$ and $t_j$)
higher than $k-1$.\footnote{This can be
seen by considering the free classical mechanics with Lagrangian
$\tr (dM/dt)^2$, and the solution $M=(1-t)M_1+t M_2$.
Changing variables to eigenvalues and unitaries, if $[M_1,M_2]\ne 0$
there will be a non-zero angular momentum $L=[M,\dot M]$, and a
potential $L_{ij}^2/(\lambda_i-\lambda_j)^2$ repelling
the trajectory from the degeneracy.}

The situation for matrices in the symmetry class CI is similar,
and governed by the same criterion \eq{linfam} applied to the
matrix \eq{deltaH}.  This implies that
a $k$-fold  eigenvalue degeneracy in codimension $k-1$.
is possible only if $Z$ is a linear sum of $k$ matrices $\hat Z_i$
which commute after applying a similarity transformation; in other
words if there exists a unitary $U$ such that
\begin{equation}\label{eq:simU}
[U \hat Z_i U^T,U \hat Z_j U^T]=0.
\end{equation}
Thus generic sums of $n$ matrices will
exhibit the generic behavior of an $n$-parameter family of matrices,
in other words eigenvalue repulsion and no $n+1$-fold degenerate
eigenvalues.

The most obvious way to get the structure \eq{linfam} and independent
$F$ breaking parameters, is the case that the effective supergravity
Lagrangian can be written as the sum of two (or more) completely
independent Lagrangians, each depending on a distinct subset of the
fields.  This fits with the intuition that in a model with several
decoupled hidden sectors, supersymmetry breaking in the various
sectors would be independent, but only if the different sectors
satisfy a very strong decoupling requirement.  One way to get this is
to consider limits in moduli space, and following up this idea leads
to the question, what would be the consequences of a nearly degenerate
eigenvalue.  From \eq{defdtwoV}, this would seem to lead to a nearly
massless boson, but no obvious enhancement of the number of vacua.
This case deserves closer study, but we leave this for future work.

It turns out that less simple but naturally occuring examples can also
satisfy \eq{linfam} and lead to degenerate eigenvalues; we will see
this below for the case $T^6/\IZ_2\times\IZ_2$.  However, this example
has an unusual degree of symmetry, and we have seen no sign of the
structure \eq{linfam} in other Calabi-Yau compactifications.

\subsection{Some physical comments}

While we suspect that some of these observations have been made
before, our discussion does not look much like the existing discussions
of supersymmetry breaking in the literature, which made it hard for us to
compare with previous work.  There are two reasons for this.  First,
in model building terms, the considerations here would describe the
origin of supersymmetry breaking in hidden sectors, and not its
mediation or effects on the observable sector.  Second, we have tried hard
to use the underlying geometry to simplify the problem to its essentials,
perhaps at the cost of some physical intuition.  Let us make a few comments
to remedy this.

The matrix $M$ is closely related
to the supergravity fermion mass matrix, which can be read off from
the Lagrangian
(\cite{WessBagger}, (23.3)):
$$
\CL_F = W \bar\psi_a \sigma^{ab}\bar\psi_b
 + \bar W \psi_a \sigma^{ab}\psi_b
 + \frac{i}{\sqrt{2}} F_A \chi^A \sigma^a\bar\psi_a
 + \frac{i}{\sqrt{2}} \bar F_A \bar\chi^A \sigma^a\psi_a
 + \half Z_{AB} \chi^A\chi^B
 + \half \bar Z_{AB} \bar\chi^A\bar\chi^B
$$
where $\psi^a$ is the gravitino, $\chi_A$ are the fermionic partners
of the moduli $z^Z$, and $\bar\psi^a$ and $\bar\chi_A$ are related
to these by complex conjugation. Making the change of variables
$\psi^a\rightarrow e^{i\theta/2}\psi^a$ and $\chi_A\rightarrow
e^{-i\theta/2}\chi_A$, and decomposing $\chi$ into the eigenvectors
\eq{takagidecom}, one finds that the eigenvector $\psi_F$ with
eigenvalue $2|W|$ is the goldstino, while the other eigenvectors
$\psi_a$ lead to fermions with Majorana masses $m_A=\lambda_A$.
The condition that $M$ has an eigenvalue $2|W|$ is also the same as
the condition for a scalar in a supersymmetric vacuum to go
massless, as discussed in section 3.2 of \cite{dd}.  This is not a
coincidence; it is because different physical branches of the space
of vacua attach at second order phase transitions, {\it i.e.} points
at which a scalar field becomes massless.  As we vary $N$, when its
kernel jumps, a modulus will go massless.  Note however that there
are two types of variation of $N$, those which come from varying the
moduli $z^A$, but also those which come from varying the effective
Lagrangian data $(W,F,Z)$.  While all supersymmetry breaking
branches are connected to the supersymmetric branches by varying
$N$, these will be physical second order transitions only if they
are connected by varying moduli.

One way to see the origin of the stability condition $\lambda_a \ge
\lambda_1=2|W|$ is that the bosonic mass matrix \eq{dV} has a
universal term $\delta_{AB}(|F|^2-2|W|^2)$, which for $|F|<<|W|$
is a universal tachyonic contribution.  This is not much emphasized in
the literature, which usually considers the case $D=0$ or equivalently
$|F|^2=3|W|^2$, in which case this contribution is positive.  However
it is quite important in general, especially in the case $|F| << |D|\sim |W|$.

\section{Distributions of non-supersymmetric vacua}

We proceed to
work out the distribution of supersymmetry breaking parameters,
\begin{equation} \label{eq:nonsusyint}
 d\mu[F] =
 \int d\mu[W,F,Z,U] \ \delta^{2m}(d V) \, |\det d^2 V|.
\end{equation}
We start by changing variables from $Z_{AB}$ to its eigenvalues
and eigenvectors $\lambda_A$ and $\psi_A$. This gives us a joint
distribution
 $$
d\mu[W,F,\lambda_A,\psi_A,U] .
 $$
These variables are redundant under permutations in $S_N$ acting
on the indices $A$.  To fix this, we take the $\lambda_A$ in
increasing order, as in \eq{ordering}.

We then solve the constraint $dV=0$ as in (\ref{eq:solconstr}), in
a basis with $Z$ diagonal. Put $\phi_A = \arg (e^{-i \theta/2}
F_A)$. We focus on solutions with $2|W|=\lambda_1$, because this
is where the bulk of the metastable vacua are located, as argued in
section \ref{sec:masses}. Factorizing $\delta(N\cdot F)$ (with
$N\cdot F$ as in \eq{defN}) turns
\begin{eqnarray}
d^{2m}F\ \delta^{(2m)}(N\cdot F) &=& d^{2m}F\ \prod_A
  \delta[\Re (\lambda_A \bF_A - 2|W|e^{-i\theta} F_A)] \,
  \delta[\Im (\lambda_A \bF_A - 2|W|e^{-i\theta} F_A)] \nonumber \\
&=& d^{2m}F\ \prod_A
  \delta[|F_A|\cos \phi_A (\lambda_A -2|W|)]
  \delta[|F_A| \sin \phi_A (\lambda_A  + 2|W|)] \nonumber \\
&=& d^{2m}F\ %
  \frac{1}{|F_1|}\delta(\lambda_1-2|W|) \,
  \frac{1}{|F_1|(\lambda_1+2|W|)}\delta(\sin \phi_1) \prod_{A>1}
\frac{\delta^2(F_A)}{\lambda_A^2-4|W|^2} \nonumber
\\ &=& \frac{df}{|f|} \frac{\delta(\lambda_1^2-4|W|^2)}{|\det' N|}
\label{eq:deltas}
\end{eqnarray}
where $\det' N$ is the determinant with the modes $\Psi_1^{\pm}$
excluded, and $F$ is given by (\ref{eq:solconstr}) with $A=1$.

For small $F$, the determinant of $d^2V$ can be approximated by
\begin{eqnarray} \label{eq:detddV}
\det d^2V &\approx& \prod_A (m^+_A)^2 (m^-_A)^2 \nonumber
\\
 &\approx&  (m_1^+)^2 (m_1^-)^2 \prod_{A>1} (|W|^2 - \lambda_A^2)(4|W|^2-
\lambda_A^2)
  \nonumber \\
 &=& -\frac{\det H}{3|W|^2} \, {\det{}' N} \, (m_1^+)^2 (m_1^-)^2.
\end{eqnarray}
The $m_A$ are the masses introduced in \eq{masses}, and we assumed
the $\lambda_A$ for $A>1$ to be sufficiently above
$\lambda_1=2|W|$ such that the higher order terms can be neglected
in these $m_A$. Thus we can extract the $\det' N$ from the
numerator, to cancel with the one in \eq{deltas}. The $1/3|W|^2$
here cancels the factor in $\det H$ coming from the
eigenvectors $\Psi^{\pm}_1$, so this expression is non-singular.

The result for the distribution of nonsupersymmetric vacua
is $d\mu[W,F,\lambda,\psi,U] \, \rho$, where
\begin{equation} \label{eq:branchone}
 \rho =
 \delta^{2m-1}(F)\,
 \delta(\lambda_1^2-4|W|^2) \,
 \frac{|\det H|}{3|W|^2} \cdot \frac{(m_1^+)^2 (m_1^-)^2}{|F|}
\end{equation}
and $\delta^{2m-1}(F)$ is interpreted as above, and
$(m_1^{\pm})^2$ is as in \eq{msq} and \eq{msqn}.

Most of the structure of the result is in the
factor $|\det H|$.  This is the same as the Jacobian appearing in the
density of supersymmetric vacua, so the result can be summarized
as saying that the density of non-supersymmetric vacua is very
similar to that for supersymmetric vacua,
with a ``correction factor'' proportional to the product of the
squared masses of the two moduli in the supersymmetry breaking
direction.

If we were on a locus with $k$ degenerate eigenvalues, we would have
to generalize the above steps: \eq{deltas} would contain a product
of $k$ delta functions $\delta(\lambda_i-2|W|)/F$, we would have a
larger kernel of $N$, and would need to replace \eq{detddV} with the
determinant of a $k\times k$ submatrix of $V''_1+V''_2$. While
possible in principle, normally this will happen only in
codimension larger than $k$, as discussed earlier, and thus will not
contribute to the final integral.

If we do not impose conditions on the value of the cosmological
constant, we have generically (i.e.\ for $|W|$ not too small)
$(m_1^-)^2 \approx 4|W|^2$ and the distribution becomes
\begin{equation}
 \rho =
 \delta^{2m-1}(F)\,
 \delta(\lambda_1^2-4|W|^2) \, \Theta_+(\theta) \,
 \frac{|\det H|}{3|W|^2} \cdot \frac{4|W|^2 |\Re ( e^{2i\theta} U_{ABC} \bF^A
\bF^B \bF^C )|}{|F|^3} .
\end{equation}
Here $\Theta_+(\theta)$ restricts $\theta$ to values for which
$(m_1^+)^2>0$ (which is approximately half its integration
domain). Thus we see that the small $|F|$ behavior of the
distribution is $d F$, so the number of generic metastable
supersymmetry breaking vacua with
supersymmetry breaking scale \eq{defMsusy} less than $M_*$
\begin{equation}
 \CN(M_\susy<M_*) \sim M_*^2
\end{equation}

If on the other hand we restrict to metastable vacua with $V \sim 0$
(and $D=0$),\footnote{By $V \sim 0$ we mean $|V|<<O(F)$, which can
still be much larger than the observed cosmological constant. In
particular further quantum corrections to $\Lambda$ remain
approximately within this window.} we have to multiply this
distribution by $\delta(|F|^2-3|W|^2)=\delta(|F|-\sqrt{3}|W|)/2|F|$,
leading to
\begin{equation}
 \rho = \delta(|F|-\sqrt{3}|W|) \,
 \delta^{2m-1}(F)\,
 \delta(\lambda_1^2-4|W|^2) \, \Theta_+(\theta,F) \,
 \frac{|\det Z|^2}{4 |W|^2} \cdot \frac{(m_1^+)^2
 (m_1^-)^2}{2|F|^2} \\
\end{equation}
where $\Theta_+(\theta,F)$ restricts $\theta$ and $F$ to values
such that $(m_1^{\pm})^2>0$, which as argued in section
\ref{sec:masses} gives at least an additional $O(F)$ suppression.
We also used $|\det H| \approx 3|W|^2 \prod_{A>1} \lambda_A^2 =
\frac{3}{4} |\det Z|^2$.

Let us now grant that the joint distribution we started with
looks like
$$
d\mu[W,F,Z,U] \sim d^2W~ d^{2n+2}F~ d\mu\ldots ;
$$
in other words $W$ and each component $F_A$ are roughly uniformly
distributed in the complex plane,
and independent of each other and the $Z$'s.
All this is true in the \IIb\ case.
We can then use the delta functions to solve for $W$ and $|F|$,
to obtain
\begin{equation} \label{eq:lowWdist}
\rho \sim d\mu[\lambda,U]
dF\, \Theta_+(\theta,F) \,
 \frac{|\det Z|^2}{4 |W|^2} \cdot \frac{(m_1^+)^2
 (m_1^-)^2}{2|F|^2} .
\end{equation}
As discussed in section \ref{sec:masses}, metastability forces
$(m_1^+)^2 \sim (m_1^-)^2 \sim F^2$ and $\Theta_+(\theta,F) \sim
F^2$, leading to
$$
\rho \sim d\mu[\lambda] F^4 dF .
$$
Furthermore, in a generic distribution of superpotentials such as
\eq{CIdist}, one expects
$$ d\mu[\lambda] \sim \lambda d\lambda \sim F dF ,
$$
leading to
$$
\rho \sim F^5 dF .
$$

Thus, the number of metastable susy breaking vacua with near-zero
cosmological constant and susy breaking scale $M_{\susy}<M_*$ goes like
\begin{equation} \label{eq:tenthpower}
  \CN(M_{\susy}<M_*,\Lambda \sim 0) \sim M_*^{12}.
\end{equation}
Among such vacua, small supersymmetry breaking scales are
disfavored. Note also that this result suggests that metastable de
Sitter vacua obtained by pure F-term susy breaking are relatively
rare.

\subsection{Type IIB flux vacua}

For type IIB flux vacua (ignoring K\"ahler moduli), we have
\cite{dd}:
\begin{equation}
 Z_{00} = 0, \qquad Z_{0I} \equiv Z_I, \qquad Z_{IJ} = \CF_{IJK}
 \bZ^K
\end{equation}
and
\begin{equation}
 U_{00I}=0, \qquad U_{0IJ}=\CF_{IJK} \bF^K, \qquad U_{IJK}=D_I
 \CF_{JKL} \bZ^L + \CF_{IJK} \bF^0.
\end{equation}
The nonzero curvature components are
\begin{equation}
 R_{0\bar{0}0\bar{0}} = -2, \qquad R_{I\bar{J}K\bar{L}} =
 \CF_{IKM}{\bar{\CF}^M}{}_{JL} - \delta_{IJ}\delta_{KL} - \delta_{IL}\delta_{JK}.
\end{equation}
This gives
\begin{eqnarray}
 (m_1^\pm)^2 &=& \frac{2}{|F|^2} \left( |\CF_{IJK} \bF^J \bF^K \pm 2 e^{-2 i
\theta} F_0 F_I|^2 - 2|F|^4
 \pm \re(e^{2i\theta} D_I \CF_{JKL} \bZ^I \bF^J \bF^K \bF^L)
 \right) \nonumber \\ \label{eq:msqflux}
 && + \delta_{\pm,-} \, 4 |W|^2,
\end{eqnarray}
so indeed when $F \sim W$ and $D\CF \sim O(1)$, to keep both
squared masses positive, one needs an order $F$ fine-tuning of the
term proportional to $D\CF$. At large complex structure, $D\CF=0$
so no such fine tuning is needed.\footnote{However, the remaining
terms may still fail to be positive.} On the other hand, if the
number of moduli is large, this constitutes only a tiny part of
the moduli space and hence of the number of vacua. Near a conifold
degeneration, both $D\CF$ and $\CF$ blow up, with $D\CF \sim
\CF^2$, so the same kind of fine-tuning is needed as in the
generic case.

\subsubsection{One complex structure modulus}

We turn to some more explicit results, beginning with the simplest
case of one complex structure modulus. We will see that this
illustrates some but not all features of the generic discussion above.

Letting $Z\equiv Z_{01}$, the matrix $Z_{IJ}=D_I D_JW$ is
$$
Z_{IJ} = \left(\begin{array}{cc}
0& Z\\ Z& \CF \bZ
\end{array}\right) ,
$$
and the eigenvalues $\lambda$ of $M$ satisfy
$$
(\lambda^2-|Z|^2)(\lambda^2-|Z|^2(1+|\CF|^2)) - |\CF|^2|Z|^4 = 0.
$$
More explicitly, $\lambda_{1,2} = \half (|\CF| \pm \sqrt{|\CF|^2+4})
|Z|$. Near a conifold limit $\CF$ diverges, and $\lambda_1 \sim
|Z|/|\CF|$, $\lambda_2 \sim |\CF| |Z|$. The measure for $Z$ in this
limit is $d|Z|^2 = |\CF|^2 d\lambda_1^2$.

As in the general discussion, we consider $|W|,|F| << 1$. Here this
requires taking $|\CF|>>1$. The constraint in \eq{IIbdist} forces
$|Z|^2\sim L$, so we write $Z=\sqrt{L}e^{i\phi}$. Then, at a
metastable pure F-breaking critical point of $V$, $|F|^2/3 \sim
|W|^2 = \lambda_1^2/4 \sim L/4 |\CF|^2$, which is indeed small when
$\CF \to \infty$. Granting $m_{\pm}^2\sim F^2$ and substituting into
\eq{lowWdist}, we find
\begin{eqnarray*}
d\mu[F] &= \int d^2W~d^2Z~dF~ d\theta\ \delta(|Z|^2-L)
 \delta(|F|-\sqrt{3}|W|)\delta(\lambda_1^2-4|W|^2)
\Theta_+(\theta,F) \,
 \frac{|Z|^4}{4 |W|^2} \cdot \frac{(m_1^+)^2
 (m_1^-)^2}{2|F|^2} \\
&\sim \frac{3}{8} L^2~ d\phi~ d\theta~ \Theta_+(\theta,F) \times
dF~ \delta(F-\frac{\sqrt{3}}{2|\CF|}\sqrt{L}) .
\end{eqnarray*}
Granting the behavior $\Theta_+ \sim |F|^2$, this looks like a
scaling $F^2 dF$ at a given point in moduli space. What happened to
the $|F|^5 dF$ we found earlier?  One factor $|F|^2$ was cancelled
by the $|W|^2$ in the denominator -- the general expectation that
$|\det Z|^2$ would go as $|W|^2$ is violated in this case, because
the see-saw mechanism pairs a large eigenvalue with the small
eigenvalue. The remaining missing factor of $F$ comes from the fact
that in this particular ensemble, there is no additional suppression
of small $\lambda$ from the $Z$-measure, as $Z$ does need to be
tuned small to make $\lambda$ small.

Finally, we need to incorporate the dependence of $\CF$ on moduli.
Suppose we are near a conifold limit, parameterized by a single
complex structure modulus $t\rightarrow 0$; then
\begin{equation}\label{eq:Fconi}
|F|^2 \sim |W|^2 \sim \lambda^2 \sim L|\CF|^{-2} \sim L|t|^2 \log^3 |t|^2
\end{equation}
and the measure on moduli space $d^2z \sqrt{\det g}$ becomes
$$
d^2t~ \log |t|^2 \sim \frac{d^2F}{L \log^2 |F|^2}
$$
leading to
$$
d\mu[F] \sim L \frac{d^2F}{\log^2 |F|^2} \Theta_+ .
$$

The final distribution (with the metastability constraint leading to
$\Theta_+ \sim F^2$) goes as $F^3 dF$.  The difference with our
previous general claim $F^5 dF$ arises from the see-saw factor
$1/|F|^2$ in $|\det Z|^2$ (required by the scaling $\det Z\sim
F^0$).

Thus, despite the fact that flux vacua are dual to gauge theory in
the conifold limit, and do lead to hierarchically small scales,
the final vacuum distribution does not show a corresponding
enhancement at small scales.  This came from a combination of effects,
which we summarize again here and in the conclusions.

The hierarchically small scale is $|t|$, and it is true (\eq{Fconi})
that this leads to $W \sim t$, which after imposing $\Lambda=0$
implies $F\sim t$.  On the other hand, the enhancement of the number
of vacua near the conifold point found in \cite{dd} is not present
for the small $W$ component, in present terms because the large
matrix element $\CF$ cancels out of $\det Z$. In addition, the
metastability condition \eq{metaconst} gives an extra $F^2$.

\subsubsection{A simple but nongeneric multiparameter example: $T^6/\IZ_2^2$}

Another example which can be treated exactly
is the orbifold $(T^2)^3/\IZ_2^2$ (without discrete
torsion). This has three complex structure moduli $\tau_i$, and the
only nonzero $\CF_{IJK}$ is $\CF_{123}=1$. As usual we ignore the
K\"ahler moduli. The $Z$-matrix is
 $$
  Z = \left(
  \begin{array}{cccc}
   0 & Z_1 & Z_2 & Z_3 \\
   Z_1 & 0 & \bZ_3 & \bZ_2 \\
   Z_2 & \bZ_3 & 0 & \bZ_1 \\
   Z_3 & \bZ_2 & \bZ_1 & 0
  \end{array}
  \right)
 $$
Its eigenvalues and eigenvectors can be obtained by elementary means.
The eigenvalues are
 $$
  \lambda_1=|Z_1|+|Z_2|-|Z_3|, \quad
  \lambda_2=|Z_1|-|Z_2|+|Z_3|, \quad
  \lambda_3=-|Z_1|+|Z_2|+|Z_3|, \quad
  \lambda_4=|Z_1|+|Z_2|+|Z_3|,
 $$
and the eigenvectors $\psi_A$ have simple expressions which depend
only on the phases $\phi_i$ of $Z_i$:
 $$
  (\psi_A)^B =
  \frac{i}{2} \left(
  \begin{array}{cccc}
   1 & -1 & -1 & 1 \\
   1 & -1 & 1 & -1 \\
   1 & 1 & -1 & -1 \\
   i & i & i & i
  \end{array}
  \right)
  \left(
  \begin{array}{cccc}
  e^{i(\phi_1+\phi_2+\phi_3)/2}&0&0&0\\
  0& e^{i(\phi_1-\phi_2-\phi_3)/2}&0&0\\
  0&0& e^{i(-\phi_1+\phi_2-\phi_3)/2}&0\\
  0&0&0& e^{i(-\phi_1-\phi_2+\phi_3)/2}
  \end{array}
  \right).
 $$

Note that since the $Z$-measure is just $d^6Z$, there is no
suppression of coincident or zero eigenvalues.
This can be understood along the lines of the
discussion at the end of \ref{sec:degeneracies}, by noting that the
matrices $\hat Z_i = \partial Z/\partial |Z_i|$
satisfy \eq{simU} with $U_A^B=(\psi_A)^B$ independent of $|Z_i|$,
and thus can be simultaneously
diagonalized on a three real dimensional slice of parameter space.

This feature will be destroyed by generic perturbations of the
prepotential.  A simple example for which this is easily verified is
obtained by adding ``conifold-like'' terms
$\CF_{IJK}=\delta_{IJ}\delta_{IK}/t_K$.
Thus, we believe the present example is not typical, and
that more generic flux ensembles will share the CI ensemble behavior.
However, let us finish the discussion for completeness.

Putting $F=\psi_1$, \eq{msqflux} gives
 $$
  (m_1^+)^2 = 2 |F|^2, \qquad (m_1^-)^2 = 4(|W|^2-|F|^2).
 $$
This simplification and in particular independence of $\theta$ and
vanishing of the $O(F)$ term occurs because $D \CF = 0$ in this
case. This is another nongeneric feature of this model. One
consequence is that there is less tuning control of the masses, and
in particular in the case of pure F-breaking and approximately
vanishing cosmological constant, $(m_1^-)^2=-8|F|^2/3<0$, so there
are no such metastable vacua. When a constant D-term is added, this
becomes $(m_1^-)^2=4(D^2-2|F|^2)/3$, which is positive for
sufficiently small $F$.

Plugging this in the general formulae, we find for the vacuum
density at zero cosmological constant
\begin{eqnarray*} 
 d\mu &\sim& d\Lambda \,
 d^2W \, d^6 Z \, dF \, \delta(|F|^2+D^2-3|W|^2) \,
 \delta(\lambda_1^2 - 4|W|^2) \\
&& \times \Theta(|W|-|F|) \cdot (\lambda_2 \lambda_3
 \lambda_4)^2 \cdot \mbox{$ \frac{(m_1^+)^2 (m_1^-)^2}{|F|}$} \\
 &\sim& d\Lambda \, dF \, {|F|(D^2-2|F|^2) \over \sqrt{|F|^2+D^2}} L_*^{11/2} .
\end{eqnarray*}
The exponent of $L_*$ can be checked by counting dimensions, where
$F,W,Z$ get assigned dimension 1, and thus $L,\Lambda$ get dimension
2. The total dimension must be $2 b_3 = 16$, which is indeed the
case with the above power of $L_*$.

When $F << D \sim W$, this goes as $dF F$, which is as in the
generic case but with an extra factor of $F$, due to the fact that
$(m^+_1)^2$ is of order $F^2$ here. When $F \sim D$ (but not too
close to $D/\sqrt{2}$), this goes as $dF F^2$. This has two powers
less than the generic case, one because there is no additional
suppression here of small $\lambda$, and one because no fine-tuning
is needed to kill the $\pm O(F)$ term in the masses. Finally, when
$|F| \to D/\sqrt{2}$, the density drops to zero together with the
mass $m_1^-$.

Finally, there would also be terms from the branches with degenerate
eigenvalues, of the same form multiplied by powers of $F/\sqrt{L}$.
If we trust the results all the way to $|F|\sim \sqrt{L}M_{string}^2$,
which might be reasonable for $M_{string}<<M_p$ (as at very weak
string coupling), these will be comparable in number to the vacua we
just described; however as we discussed their existence appears to
be a special feature of this example.

\subsubsection{The generic multiparameter case}

We now discuss what we would expect to see for a more generic
multiparameter case, in the continuous flux approximation.  To get low
scale supersymmetry breaking with zero c.c., we need small $W$ amd
$F$.  While in this approximation, there is no {\it a priori}
constraint on the size of these parameters, solving the equations
\eq{defN} requires the matrix $Z$ to have a small eigenvalue.

It is hard to tune a generic matrix to get a small eigenvalue. This
is easily made precise in the standard ensembles of random matrices.
In the CI ensemble, the measure for matrices with a small eigenvalue
$\lambda$ goes as $d(\lambda^2)$.  Following this through leads to
the generic $F^5 dF$ distribution we discussed above.

In the one parameter case, $Z$ is a $2\times 2$ matrix, and the
dilaton dependence of the \IIb\ flux superpotential leads to a see-saw
structure.  This is special to $2\times 2$ matrices.  More generally,
what we can accomplish by tuning to conifold points is to get large
matrix elements in $Z$, for example the $\CF Z\sim Z/t$.
In higher dimensions, large matrix elements do not generally lead to small
eigenvalues.

A simple model which may illustrate the general situation is to
take $Z$ to have generic order $1$ coefficients except for a single
large matrix element $Z_{11} = \CF >> 1$.  The eigenvalues of such
a matrix can be found by treating the $O(1)$ coefficients as a perturbation
around the spectrum of the matrix for which only $Z_{11}=\CF$ is
non-zero.  A generic $O(1)$ perturbation will shift these eigenvalues
by $O(1)$, and the resulting eigenvalues will be $\CF+O(1)$ and
the rest $O(1)$.

The see-saw matrix escapes this general result because $Z_{00}=0$
forces $\det Z\sim 1$ and thus a small eigenvalue must appear.  To
get this effect in a higher dimensional matrix, the determinant of
the $1,1$ minor of $Z$ must vanish.  This would appear to be a rather
complicated and non-generic tuning.

Thus, we see no clear way out of the generic $F^5 dF$ prediction of
our earlier discussion in this case.  Since even in the one
parameter case, the expected enhancement of low scale vacua was
cancelled by measure factors, the suppression of low scale vacua of
this type would appear quite general.

\subsection{Quantized fluxes}

As in \cite{ad,dd}, we have ignored flux quantization in the results
so far, instead taking the fluxes to be continuous variables.  At
first sight this may seem a drastic simplification as the allowed
values of the EFT parameters $(W,F,Z)$ are heavily influenced by
flux quantization.  For example, the hierarchically small value of
certain contributions to the superpotential near the conifold point
is only apparent in the quantized flux problem.

Now, as in \cite{ad,dd}, one can argue on general grounds that the
approximation of taking flux continuous reproduces the leading large
$L$ asymptotics for the number of vacua, because the volume of the
region in flux space supporting vacua is the asymptotic for the number
of quantized flux vacua (lattice points contained in the region).
We make this argument below.
Furthermore, explicit numerical study
\cite{Giryavets:2004zr,dd,Conlon:2004ds,DeWolfe:2004ns}
has confirmed the validity of
the approximation in counting supersymmetric vacua when $L
\ge K/r^2$, to estimate the number of vacua in a region of moduli
space of radius $r$, in finding the distribution of $W$, and other
observables.

This does not mean that we should immediately accept the analogous claims
for nonsupersymmetric vacua, as there are clearly important
differences between the two problems.  For example, the validity of
statistical approximations to obtain the $W$ distribution is surely
helped by the fact that $W$ is a sum of complex quantities with
arbitrary phases, which typically involves many cancellations.  On the
other hand, since the quantity $M_{\susy}^4$ is a sum of squares,
errors in the approximation might lead to systematic overestimates.

More work will be needed to find the minimal flux $L$ for which the
results here are accurate, but let us proceed to make some general
comments.

\subsubsection{Geometry in flux space}

We now describe the region in type \IIb\ flux space which supports
nonsupersymmetric vacua, in the same sense developed for
supersymmetric vacua in \cite{ad}.

We work around a point $z$ in moduli space and make the
appropriate decomposition $(W,F,Z)$ of the fluxes at that point.
We then think of the $(W,F)$ directions as fibered over the $Z$
plane.  At a given $Z$ and $\theta$, the branches of
nonsupersymmetric vacua sit at the $m$ values of $|W|$ given by
the eigenvalues of $Z$, and in one-dimensional subspaces of the
$F$ space given by the eigenvectors. These $m$ solutions of the
eigenvalue equation or sheets are the basic ``branches of
solutions'' in this problem.  Of these, only the one with the
lowest value of $|W|$ is tachyon-free.  Moving around in the $Z$
plane varies them, and one can have monodromies about points with
degenerate eigenvalues which exchange sheets.

Varying $\theta$ then fills out a correlated circle in $W$ and
$F$, to give two real dimensional sheets in the $2m+2$ dimensional
$(W,F)$ space. Finally, varying the $2m$ moduli will turn these
sheets into cones or balls of full dimension.

Another way to think about this is to consider the ``universal''
solution space as $(W,F)$ fibered over the entire $m(m+1)$ real
dimensional space of $Z$, which contains a real codimension $2m$
domain of allowed nonsupersymmetric vacua (say with $|F|^2$ and
$|W|^2$ bounded) which looks like a two dimensional sheet in
$(W,F)$. Each point in moduli space then produces a rank $4m$
lattice (in the \IIb\ theory; on fourfolds it could be larger) in
this space, and then varying the $2m$ real moduli allows these to
hit the universal solution space at isolated points, the
non-supersymmetric vacua.

In any case, the region in flux space which supports nonsupersymmetric
vacua is of full dimension in the flux space and has a smooth
boundary, and thus standard arguments imply that the leading large $L$
asymptotic for the number of lattice points contained in this region
will be its volume.

\subsubsection{Subleading components in $L$}

Perhaps a subleading term at large $L$ has a different $F$ distribution,
for example producing many more small $F$ vacua, in a way which makes
it dominate in the physical regime.

While the true distribution of $F$ is quantized, one would expect this
to cut out small $F$ vacua.  In particular, one might naively expect
the distribution at a value $F$ to be well approximated by the large
$L$ asymptotic only when $L > 1/|F|^2$.  On the other hand, small
parameters enter into this relation, as we saw in the explicit one
parameter example, so the situation in cases of interest is probably
better than this.  In any case, this effect will shift the
$F$ distribution towards higher scales.

On the other hand, it has been suggested by Dine {\it et al} \cite{dgt} that
the $W$ distribution could obtain a component highly peaked at zero,
which after tuning the c.c. would lead to a peak in $F$.  Their idea is that,
because $W=0$ restores R symmetry, there might be an enhanced number
of flux vacua with $W=0$.  Then, since our exact considerations are
of course just approximations to the full physical problem,
one might expect such a peak to be smoothed out to an enhancement
of small $|W|$ vacua in the full theory.

Recently DeWolfe {\it et al} \cite{DeWolfe:2004ns}
have studied the problem of finding
supersymmetric flux vacua with $W=0$ in some detail
and indeed find enhanced numbers of $W=0$ vacua
at subleading order in $L$ in simple examples.  While it remains to be
seen how important this effect is, the idea and evidence for it seem
quite reasonable.  However, it seems unlikely that this comes with an
enhancement of small $W$ vacua in the pure flux vacuum problem.
This is for both mathematical reasons (the points with an enhanced number
of vacua tend to be surrounded by ``voids'' without vacua), and
physical reasons -- while the flux vacua do contain exponentially
small effects, the superpotentials which stabilize all moduli also contain
other $O(1)$ contributions which show up in $W$ and
cannot be eliminated.

Rather, one needs to call on additional physical effects to smooth out
the distribution.  At first sight the suggestion of \cite{dgt}
seems plausible: exponentially small corrections could lead to vacua
with small $W \sim \exp -1/g^2$, and a uniform distribution of
couplings $dg$ would translate into a logarithmic distribution
$dW/W$.  However, at present we know of no explicit ensemble of
models in which this idea would be realized.  In particular, there is
no evidence that the KKLT construction \cite{KKLT} would do this.
While it relies on exponentially
small effects to stabilize K\"ahler moduli, these are balanced against
a preexisting small $W$ obtained from the flux sector, and the
construction does not work without this.  One might hope that a model
more like the original ``racetrack,'' in which several competing
exponentials stabilize the K\"ahler moduli, could lead to small $W$, but
if the potential is naturally a polynomial in the exponentials
$q_i\equiv \exp -1/g_i^2$, one might expect the resulting distributions
to be uniformly distributed in the variables $q_i$.
All this is not to say
the suggestion is clearly false, but rather that to support it
one needs to show that it is realized
in some explicit ensemble of models which could
plausibly come out of string compactification.

In our opinion, at present the best motivated conjecture one could
make for such an ensemble of K\"ahler stabilized models, is simply
that it is similar to one of the known flux ensembles.  As in
\cite{stat}, one might try to argue this from the existence of
gauge-flux dualities such as \cite{Gop-Vafa} which (in much simpler
examples) explicitly relate the two classes of vacua.  Not having a
compelling argument of this type, we will simply make the comment that
if the true ensemble of K\"ahler stabilized \IIb\ models turns out to
be different, one will also want to understand why this type of
duality argument fails.

Suppose we were to grant that there is a large component of vacua
with small $W$ distributed as $d^2W/|W|^2$; how would this influence
our results?  We need to first ask if this component is already
visible in our computations, and properly taken into account.
Although there are similar looking factors in our intermediate
steps, in fact they are not present in the final results, for the
reasons we explained.  Rather, we would interpret the suggestion of
\cite{dgt} as saying that considerations in a sector of the theory
not considered here lead to an additional factor $|W|^{-2}$ in the
original distribution \eq{IIbdist}.  If we incorporate such a
factor, given $F\sim W$ we will find the generic distribution
changes from $F^5 dF$ to $F^3 dF$.  This still appears to favor the
high scale, at least for the purely F breaking vacua.

\section{Including D-terms}

At present there is no explicit
ensemble of EFT's with D terms which looks simple
and universal enough to justify a detailed study.  Thus in this section
we simply add a generic additional contribution to the potential,
$$
V = |F|^2 - 3|W|^2 + \epsilon V_D,
$$
and discuss its consequences.

To get small supersymmetry breaking, we take $\epsilon$ to be
small. This would be the case for instance for a contribution
arising from an anti-D3 brane placed at the bottom of the
warped throat developing near a conifold degeneration
\cite{GKP}.
In this case, which will be our concrete example below,
$\epsilon \sim |v|^{4/3}$, where $v$ is the
complex structure coordinate given by the period of the vanishing
3-cycle.
\footnote{
While it has been argued
\cite{Dbranes,Binetruy:2004hh}
that if the anti-D3 contribution has an EFT
description, this must be a D term, we know of no description
within the usual rules of $\CN=1$ supergravity
which satisfies all known properties of the string theory construction.
However this point will not be crucial for what
follows.}

The condition for a critical point $\nabla V=0$ becomes
 $$
 N \left( \begin{array}{c} e^{-i\theta} F \\ e^{i\theta} \bF \end{array}
 \right)= - \nabla(\epsilon V^D) \equiv -\epsilon d
 $$
with $N$ as in \eq{defN}. This is solved by
\begin{equation}
 \left( \begin{array}{c} e^{-i\theta} F \\ e^{i\theta} \bF \end{array}
 \right) = -\epsilon N^{-1} d.
\end{equation}
Decomposing $d=d_A^+ \Psi_A^+ + d_A^- \Psi_A^-$ with $d_A^\pm \in
\IR$ and $\Psi_A^{\pm}$ as defined in \eq{genpsi}, we have
\begin{equation} \label{eq:Ffd}
 |F|^2 = \epsilon^2 \sum_A \frac{(d_A^+)^2}{(\lambda_A-2|W|)^2} +
 \frac{(d_A^-)^2}{(\lambda_A+2|W|)^2}.
\end{equation}

\subsection{Lifted susy vacua}

Generically (i.e.\ for $\lambda_A$ not too close to $2|W|$), we
can assume all terms in the sum to be at most of order 1. This
assumption needs some discussion for the antibrane example
because of the dependence of
$\epsilon$ on the modulus $v$, which implies $d_v \sim 1/v$.
However, the matrix element $Z_{vv} \approx \CF_{vvv} \bZ^v \sim
1/v$ in $N$ compensates for this. In other words, there will be a
pair of eigenvectors $\Psi_v^{\pm}$ approximately associated to
the $v$-direction, and although in this direction $d_v \sim 1/v$,
we also have $\lambda_v \sim 1/v$, and the two cancel against each
other in \eq{Ffd}.\footnote{We should actually go to an
orthonormal frame first, but since the metric $ds^2 \sim
\log|v|^{-2} dv d\bar{v}$ in the $v$ direction, this only induces
logarithmic corrections.} Thus, $|F|$ is generically of order
$\epsilon$.

In this case, the D-term dominates over the $|F|^2$-term in the
potential:
 $$
  V \approx -3|W|^2 + \epsilon V_D,
 $$
and the measure becomes
 \begin{eqnarray}
   \delta^{2m}(dV) |\det V''| &=& \frac{\delta^{2m}(F_A + \epsilon \, e^{i
\theta}
   (N^{-1} d)_A)}{|\det N|} \, |\det HN({\bf
   1}+(HN)^{-1}(V''_1+V''_2+\epsilon V_D''))| \nonumber \\
   &\approx&\delta^{2m}(F_A + \epsilon \, e^{i \theta}
   (N^{-1} d)_A) \, |\det H|.
 \end{eqnarray}
Again some discussion is needed to justify dropping the $V_D''$
term, since it blows up as $1/v^2$ in the $v$-direction. However in this
direction $HN \sim \lambda_v^2 \sim 1/v^2$ as well, so this cancels out.

The above expression is exactly the density for the supersymmetric branch.
Similarly, the condition for the absence of negative modes of
$V''$ is the same as for the superymmetric case: $\lambda_A >
2|W|$, up to possible corrections from $V_D''$, but if
$\lambda_A-2|W|>>\epsilon$ these corrections will be small.

Thus, these nonsupersymmetric vacua correspond to
supersymmetric vacua ``lifted'' by the D-term. The number density
of such vacua at cosmological constant $\Lambda$
will approximately be equal to the
number density of supersymmetric vacua at susy cosmological
constant $\Lambda - \epsilon V_D$. Since the latter is
nonvanishing at zero \cite{dd}, we have that for small
$\Lambda<\epsilon V_D$, this
density is essentially independent of $\Lambda$, and
roughly equal to $1/\CN_{tot}(\CR)$ when
integrated over a region $\CR$ in moduli space.

The supersymmetry breaking scale
$M_\susy$ for these vacua will be of order $\sqrt{\epsilon}$, and thus
the expected number of these ``lifted susy'' vacua with $M_\susy<M_*$
is, using the results of \cite{dd} for the susy vacuum
distribution near the conifold,
\begin{equation} \label{eq:dcondist}
 \CN(M_\susy<M_*) \sim \frac{1}{\log M_*}.
\end{equation}
That is, taking into account tuning of the Higgs mass, for this
family of near-conifold lifted susy vacua, a low susy breaking
scale is favored.

\subsection{Perturbed F-term susy breaking vacua}

The situation changes when one of the eigenvalues $\lambda_A$
approaches $2|W|$. Since we turned on the D-term potential as a
small perturbation, we still expect that $2|W|$ will be the
generic approximate lower bound on the $\lambda_A$ to get a metastable
minimum. Let us therefore assume that all $\lambda_A$ are well
above $2|W|$ except possibly $\lambda_1$. Define $u \equiv
\lambda_1 - 2|W|$. When $u \ll 1$, we have
\begin{equation}
 F \sim {\epsilon \, d_1^+ \over u} \sim \frac{\epsilon}{u}.
\end{equation}
Note that since we are interested in $F\ll 1$, we require
$u \gg \epsilon$. All $(m_A^{\pm})^2$ for $A>1$ as defined in \eq{masses}
will still be given approximately by their susy values. On the other
hand
\begin{eqnarray}
 (m_1^+)^2 &=& (u+3|W|)u + c_1 \frac{\epsilon}{u} + c_2 \frac{\epsilon^2}{u^2} +
c_3^+ \epsilon \label{eq:m1pD} \\
 (m_1^-)^2 &=& (u+|W|)(u+4|W|) - c_1 \frac{\epsilon}{u} + c_2
\frac{\epsilon^2}{u^2} + c_3^- \epsilon, \label{eq:m1mD}
\end{eqnarray}
where the $c_i$ are generically of order 1. We consider three regimes:

\vspace{.1cm}
\noindent {\it (1)} $\epsilon^{1/3} \ll u \ll 1$:
\vspace{.1cm}

In this case, the first term in \eq{m1pD} is much bigger than $\epsilon^{2/3}$,
and the remaining terms
are all much less than $\epsilon^{2/3}$. The same is true for \eq{m1mD}.
Therefore we can drop all correction terms and we are back in the previous case
of
D-term lifted supersymmetric vacua.

\vspace{.1cm}
\noindent {\it (2)} $ \epsilon \ll u \ll \epsilon^{1/2}$
\vspace{.1cm}

When we don't impose constraints on the cosmological constant, so we can assume
$|W|$ to be generic, we have that the first term in \eq{m1pD} is much smaller
than
$\epsilon^{1/2}$, while $c_1 \epsilon/u \gg \epsilon^{1/2}$. The other terms
are again much smaller, so $m_1^+$ is dominated by the second term. Furthermore
$(m_1^-)^2 \approx 4|W|^2$.
This is as in the case of pure F-term susy breaking.

The case with cosmological constant constrained near zero is similar. Since
$\epsilon^2/u^2 \gg \epsilon$, the
$|F|^2$-part of the potential will dominate over the D-term part, and thus to
get near zero vacuum energy
we should take $|W| \sim F \sim \epsilon/u \gg u$. This implies that the first
term in \eq{m1pD} is of order
$\epsilon$ and can be dropped together with the D-term, bringing us again to the
situation
of pure F-breaking. For $m_1^-$ similar considerations hold.

Hence for either case the measure $\delta^{2m}(dV) |\det V''|$ becomes
\begin{eqnarray}
 &&\delta^{2m}(F+\epsilon N^{-1}d) \, \prod_{A>1} (\lambda_A^2-|W|^2) \cdot
\frac{(m_1^+)^2 (m_1^-)^2}{u(u+4|W|)} \\
 &\approx& \delta^{2m}(F+\epsilon N^{-1}d) \, \frac{|\det H|}{3|W|^2} \cdot
\frac{(m_1^+)^2 (m_1^-)^2}{4|W| u}
\end{eqnarray}
with $m_1^{\pm}$ approximated as in the pure F-breaking case. Note furthermore
that
the delta-function forces $F$ to lie approximately in the direction of the
eigenvector
$\Psi_1^+$ as $u=\lambda_1-2|W|$ is small. Finally, since $F \sim \epsilon/u$,
we can write
 $$
  \frac{d\lambda_1}{u} = \frac{du}{u} = \frac{dF}{F}
 $$
and change variables from $\lambda_1$ to $F$ (mapping the domain
$2|W|+\epsilon \ll \lambda_1 \ll 2|W|+\epsilon^{1/2}$
to $\epsilon^{1/2} \ll F \ll 1$). This completely reproduces the measure
\eq{branchone} of pure F-breaking vacua in this regime. Thus, pure
F-breaking vacua with susy breaking scale
$M_\susy=F \gg \epsilon^{1/2}$ are just
slightly
perturbed by adding a D-term potential $\epsilon V_D$, and their distributions
remain
essentially identical.

Note that because of the suppression by at least $F_*$, these vacua can be
expected to be significantly
less numerous than the lifted supersymmetric ones.

\vspace{.1cm}
\noindent {\it (3)} $\epsilon^{1/2} < u < \epsilon^{1/3}$:
\vspace{.1cm}

This is the intermediate regime, where D-term and F-term effects
are of comparable size. When $\epsilon$ is small, this corresponds
to only a small fraction of susy breaking vacua, compared to the
other two regimes.

\subsection{Summary}

Adding a D-term which becomes small in a region $\CR$ of moduli space
will remove all vacua located outside that region from the ensemble of
vacua with ``small'' susy breaking scale ($M_\susy \ll m_p$), simply
because the D-term either destabilizes the vacuum, or it causes the
susy breaking scale to be too large. Within $\CR$, it will add new
susy breaking vacua obtained by lifting susy vacua.

For the particular case of susy breaking by an anti D3-brane near a conifold
point, the number of vacua with susy breaking
scale $M_\susy<M_*$ will approximately be given by
\begin{equation}
 \CN_0(M_\susy<M_*) \sim \frac{\CN_{susy}(\CR)}{\log M_*}
\end{equation}
where $\CN_{susy}(\CR)$ is the number of tachyon-free susy vacua in
$\CR$. Restricting the cosmological constant to a small interval of
width $\Delta \Lambda$ around zero just multiplies this number by
$\Delta \Lambda$. Apart from these lifted susy vacua, descendants of
pure F-term breaking vacua are also present and are only slightly
perturbed as long as $|F|^2$ is bigger than the added D-term
potential. Their number can be expected to be roughly
\begin{equation}
 \CN_F(M_\susy<M_*) \sim M_*^2 \, \CN_{susy}(\CR)
\end{equation}
without constraint on $\Lambda$, and
\begin{equation}
 \CN'_F(M_\susy<M_*) \sim M_*^{12} \, \Delta \Lambda  \, \CN_{susy}(\CR)
\end{equation}
if $\Lambda$ is constrained near zero.

Finally, although detailed considerations in the observable sector are beyond
our scope here, one effect which must be mentioned is the
universal $-2|W|^2$ contribution to the bosonic mass matrix \eq{DcDV},
which will destabilize vacua with light fermions and $|D|,|W|>>|F|$,
thus removing a large class of potentially realistic vacua.

\section{Conclusions}

We derived formulae for the distribution of nonsupersymmetric
vacua in a general ensemble of effective supergravity theories, such
as \eq{branchone}.
This distribution is rather similar to that for supersymmetric vacua,
with certain ``correction factors'' which we explained.

We began by reformulating the problem of finding nonsupersymmetric F
breaking vacua as that of finding eigenvectors of the matrix $D^2W$ of
second derivatives of the superpotential.  This makes several features
of the problem manifest, most importantly that metastable
nonsupersymmetric vacua are generic, of number comparable to the number
of supersymmetric vacua.

We then argued that the suggestion of \cite{dd,mrd-susy} for a large
power law growth in the number of F breaking vacua in models with many
moduli is not expected in general.  While the heuristic argument
leading to this suggestion seems sensible, namely that independent
supersymmetry breakings in independent hidden sectors combine
additively and favor high scale breaking, our detailed analysis shows
that the different F terms will be independently distributed only when
the different hidden sectors are totally decoupled, which seems
unlikely to us based on our studies so far.  It should be said that
this does not address the similar argument in \cite{suss-susy} in
terms of susy breaking by multiple antibranes, and the related idea in
\cite{mrd-susy} that multiple independent D terms could lead to the
same effect, as the details there are quite different, and this
possibility remains open.

We went on to compute the density of nonsupersymmetric vacua,
\eq{nonsusyint}, in general and in simple examples.  Much of the
structure of the result comes from the factor $\det V''$ which is
included to normalize the delta function $\delta(V')$ to one for each
vacuum.  Physically, this factor is the product of masses squared for
all bosons, and this makes precise the general expectation that
bosons much lighter than the natural scale of the potential (for
flux potentials, the string scale) are disfavored.

Perhaps the simplest summary of the results for pure F breaking is
\eq{tenthpower}, which states that the number $\CN$ of pure F
breaking vacua with supersymmetry breaking scale $M_{\susy}\le M_*
<< M_{pl}$, and with $\Lambda\sim 0$, goes as $\CN \sim M_*^{12}$.
This somewhat surprising claim is made up of a number of factors,
explained in detail in section 3:
\begin{itemize}
\item The most naive expectation would have been a uniform
distribution $d^2F$ for the complex parameter $F$, leading to
$\CN\sim M_*^4$.
However, as discussed in section 2.2, the equations $V'=0$ determine
the phase of $F$, leading instead to the generic
distribution $dF$ and the counting $\CN\sim M_*^2$,
for vacua in which the c.c. is not tuned, or in which
it is tuned by other effects (say by D terms).
\item If we tune the c.c. to zero using $|F|^2=3|W|^2$, we need to
take the $W$ distribution into account.  $W$ is a complex variable,
and is proportional to the lowest eigenvalue $\lambda$
of the matrix $Z\equiv D^2W$ (in the sense of \eq{takagidecom}).
Since $Z$ is complex symmetric,
this is distributed as $d(\lambda^2)$, leading to an extra factor
of $\lambda\sim W\sim F\sim M_*^2$.
\item The measure factor $\det V'' = \prod_i m_i^2$ weighs the
density of vacua with the product of the masses squared of every boson.
As discussed in section 2.3, the two bosonic partners to the goldstino
generically have masses determined by $W$ and $F$,
and this leads to a factor $F^2\sim M_*^4$.
\item Finally, metastability requires a tuning
which leads to an additional factor $F^2\sim M_*^4$, as discussed in
section 2.3.
\end{itemize}
In sum, enforcing $\Lambda\sim 0$ has a significant effect on the
distribution, apparently favoring high scale breaking
enough to outweigh factors such as
\eq{hierarchy} and the suggestions discussed in section 3.2.2.
Note that the arguments only really require
$\Lambda < |F|,|W|$, which is good as any more precise requirement would
be spoiled by subsequent corrections.

We can compare this to a rough picture of the ``D breaking'' vacua.
Since the D parameters are real, and this type of supersymmetry
breaking need not come with light scalars, one does not get the measure
factors we just discussed, leading to distributions such as
\eq{dcondist} which would seem to favor low scale breaking.  On the
other hand, multiple D parameters might still lead to power law growth
in $M_\susy$, and a full analysis might suggest other measure factors, so
this remains unresolved.  We also do not know whether F or D breaking
is more common, since we have as yet no explicit ensemble which
includes both types of vacuum.

All of these considerations are of course within a subsector of the
theory and ignore many further corrections.  Nevertheless the factors
we just listed which entered in our final results appear fairly
generic, and thus we believe they will be important ingredients in all
computations of this type.  Numerically, they appear as important as
the specific physical effects focused on in previous work.

Much more detailed work in examples will be required to judge
to what extent the full picture is controlled by such generic
features.  One can take various attitudes about them -- perhaps they
are background, perhaps they are the signal.  But they appear to us
to be very basic properties of the distribution of string/M theory vacua.

Of course, we are still only talking about a small part of the full
problem of counting realistic string/M theory vacua.  One might expect
equally important selection factors to arise at subsequent stages, and
it is probably premature to read too much phenomenological
significance into our results so far.  But we would like to suggest
that analyses at this level of detail, at least for the best understood
classes of vacua (Calabi-Yau compactification of \IIb, F and heterotic
theories), for every part of the full problem, might be feasible over
the next few years, and that this would enable us to make similar
statements with some claim to significance.

\vskip 0.2in {\it Acknowledgements}

We acknowledge valuable discussions and correspondence with M.~Dine,
S.~Kachru, D.~E.~Kaplan, D.~O'Neil, Z.~Sun, S.~Thomas and J.~Wacker.
This research was supported in part by DOE grant DE-FG02-96ER40959.

\end{document}